# MoCLIM: Towards Accurate Cancer Subtyping via Multi-Omics Contrastive Learning with Omics-Inference Modeling

Ziwei Yang*
Bioinformatics Center, Kyoto University, Japan
yang.ziwei.37j@st.kyoto-u.ac.jp

Zheng Chen*
SANKEN, Osaka University, Japan
chenz@sanken.osaka-u.ac.jp

Yasuko Matsubara
SANKEN, Osaka University, Japan
yasuko@sanken.osaka-u.ac.jp

Yasushi Sakurai
SANKEN, Osaka University, Japan
yasushi@sanken.osaka-u.ac.jp

## ABSTRACT

Precision medicine fundamentally aims to establish causality between dysregulated biochemical mechanisms and cancer subtypes. Omics-based cancer subtyping has emerged as a revolutionary approach, as different level of omics records the biochemical products of multistep processes in cancers. This paper focuses on fully exploiting the potential of multi-omics data to improve cancer subtyping outcomes, and hence developed MoCLIM, a representation learning framework. MoCLIM independently extracts the informative features from distinct omics modalities. Using a unified representation informed by contrastive learning of different omics modalities, we can well-cluster the subtypes, given cancer, into a lower latent space. This contrast can be interpreted as a projection of inter-omics inference observed in biological networks. Experimental results on six cancer datasets demonstrate that our approach significantly improves data fit and subtyping performance in fewer high-dimensional cancer instances. Moreover, our framework incorporates various medical evaluations as the final component, providing high interpretability in medical analysis.

## KEYWORDS

Cancer subtypes, Multi-omics data, Contrastive learning *indicates corresponding authors.

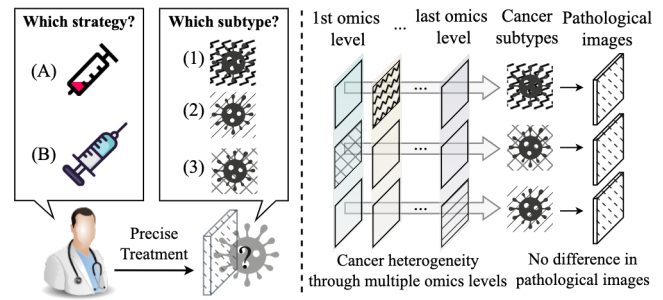

**Figure 1:** (left) There are distinct subtypes for a cancer. The clinical settings require idendifying subtypes for selecting precise treatment. (right) Cancer subtypes manifest diverse molecular information across multi-omics levels, but present similar morphological features in pathological images.

## 1 INTRODUCTION

Cancer is a pervasive public health concern that is a leading cause of death worldwide. Despite great efforts to research and treatment, the World Health Organization (WHO) reports an estimated 19.3 million new cancer cases in 2020 and 10 million cancer deaths globally [1]. The main reason for such high morbidity and mortality in cancer is its **heterogeneity**: *each specific cancer comprises multiple subtypes.* These subtypes refer to groups of patients with specific biochemical mechanisms that require tailored therapeutic strategies [2, 3]. While the subtypes may differ in their biochemical levels, they often share the same morphological traits, such as in an organism's physical structure and form [4]. This can result in high similarity in histopathological images and lead to inaccuracies in the traditional imaging tests employed in cancer diagnosis [5]. This dilemma underscores the need for continued research and new technologies to effectively identify cancer subtypes (known as ***cancer subtyping***). This not only improves clinical outcomes but also facilitates biomarker discovery for biochemical experiments.

Recent breakthrough in biological high-throughput technologies opens new possibilities for researchers to access and analyze multi-omics data from cancer patients [6]. Omics data is a comprehensive collection of molecular-level details that record disruptions or dysregulations of normal physiological processes within cancer cells [7]. Studying multi-omics data offers biochemical insights into the regulators of cellular development, which can ultimately lead to tumor formation or cancer [8] Recent works have shown promise in using omics data for cancer subtyping, e.g., by mining gene expression variations of different subtypes in transcriptomics [9–11]. This paper follows this line, and we are interested in fully exploiting the potential of omics-based subtyping. However, this task is hindered by both *data perspective* and the *underlying biological complexity*.

From the data perspective, omics data is high-dimensional, comprising tens of thousands of measurements (features) with substantial variability. In contrast, data is scarce, particularly given the ethical considerations in data collection. For example, the available cancer dataset in TCGA typically comprises hundreds of samples ($\sim 700$), yet each instance may have up to $10^6$ magnitudes in single omics data type. Also, experimental noise or technical limitations often result in a proportion of undetectable or missing values in omics data [12]. Computational analysis of such data requires feature extraction and dimensionality reduction to capture representative features of subtypes. However, conventional methods, such as principal components analysis (PCA) or Pattern Fusion Analysis (PFA), struggle to capture complex non-linear dependencies among features [13], leading to inaccurate subtyping performance. Recent booming deep learning, especially deep generative models (DGMs), shows promise in learning complex dependencies while projecting high dimensionality into a more explainable latent space (e.g., Gaussian) [14]. However, effectively learning a representative latent space is a long-standing challenge for DGM, especially when facing high-dimensional and scarce inputs [10]. The above discussion arise a question *(RQ1)* of *"How can we effectively learn a good representation of cancer omics data while avoiding data issues?"*

From a biological perspective, glancing at current omics-based subtyping works, where most efforts are made on a single-omics view [9, 11, 15]. Such methods provide partial information on the underlying pathogenesis and biomarker discovery [16, 17]. While some recent attempts have explored multi-omics, they often ignore modality-specific differences of omics, e.g., gene expression and protein, treating them uniformly [18]. Cancer omics is neither simple nor independent but interacted [19], and driven by the inference relationships of various dysregulated expressions across multi-omics [20]. Specifically, transcriptomics and epigenomics are distinct yet interrelated subsets of genomics [21]. Transcriptomics studies RNA molecules transcribed from the genome, while epigenomics examines the modifications of the genome that affect gene expression [22, 23]. Modeling such inference or biological network has long been an expectation in the computational biology to generate interpretable descriptions of behaviors [24, 25]. This is also a vital consideration when mining omics data. *"How can we effectively integrate multi-omics data, informed by biological observations?"* is the second research question *(RQ2)* in this paper.

Given these two considerations, this paper proposes MoCLIM, a representation learning framework that effectively integrates multi-omics data to improve cancer subtyping outcomes. We view multi-omics data as diverse observations derived from given cancer samples, and we independently learn omics-specific features. This approach addresses modality differences and extracts information about cancer heterogeneity from distinct omics feature spaces. Importantly, this paper positions MoCLIM in a biological axiom: genome-wide transcriptomics analysis is the mainstay of omics studies [26–28]. We incorporate multi-omics contrastive learning into our framework design. This contrasting strategy unifies the omics-specific feature spaces to maintain consistency and also can be interpreted as a projection of biological inferences between transcriptomics and other omics data. Furthermore, this paper considers a crucial question common to all biomedical studies, *(RQ3)*: *"How can we facilitate the interpretation of model-based results to support downstream biomedical analysis?"*. To this end, we present several authoritative biological metrics and provide examples of analyses informed by our learned features. This consideration further evaluates our framework as well-supported.

The contributions of this paper include the following:

- MoCLIM is a flexible framework that can effectively handle multi-omics input to generate a well-grounded representation for cancer subtyping. Superior results showed in different cancer types.
- MoCLIM models a biological observation, inter-omics inference, into a computational framework. Various ablation studies show the effectiveness and necessity of this design.
- We demonstrate how MoCLIM can support downstream biomedical analyses, generating more interpretable analytical results in cancer research.

## 2 RELATED WORK

A comprehensive understanding of human health and disease (cancer) requires the interpretation of molecular anfractuosity and variations based on multi-omics studies, such as the genome, epigenome, transcriptome, etc [29]. Multi-omics data integration and representation methods have gained great interest in studying cancer heterogeneity and subtyping [30–32].

### 2.1 Conventional Methods

Omics data is high dimensionality, scarcity, and noise. Conventional methods, such as PCA, ICA, and LASSO, are often used for preprocessing and dimensionality reduction to capture key features [33–35]. However, such methods are too Naïve, structure-based similarity metrics being a common approach, such as similarity Network Fusion (SNF) [36] and Neighborhood-based Multi-Omics clustering (NEMO) [37]. They create similarity networks to project the feature dependencies of omics data. Another strategy is distance-based metrics, such as Cancer Integration via Multi-kernel Learning (CIMLR) [38], which utilizes kernel methods to measure patient-to-patient distance and similarity matrix. Statistical methods like iClusterBayes [39], moCluster [40], and Consensus Non-negative Matrix Factorization [41] have been developed to map the data into a latent space, while identify key features via clustering like PCA.
- **Limitations.** Conventional methods are sensitive to parameter choices, and computationally intensive. They often fail to capture the complex feature dependencies effectively for data scarcity.

### 2.2 Deep Learning Methods

*2.2.1 Deep Generative Models.* Deep generative models (DGMs) have drawn recent attention as a solution for extracting complex intra-omics dependencies in unlabeled data [42, 43]. Two types of DGMS are popular: generative adversarial networks (GANs) [44–46] and Auto-encoder (AE) based models [47–49]. One early attempt is GAN-subtype, which unifies multi-omics data into a Mixture Gaussian distribution and samples a representative feature for cancer subtyping [18]. Popular AE variants, such as Variational Auto-Encoder (VAEs) [50] and XOmiVAE [51], also promise to compress high-dimensionality into an explainable latent space.

- **Limitations.** Stable GANs are different to train despite their considerable potential, particularly with few data. Moreover, the (Gaussian) assumption in both GANs and VAEs is unlikely to be fulfilled in the cancer heterogeneity, which often leads to overfitting issue [13]. While the recent work [10] leverages more flexible discrete distribution in vector-quantized VAE (VQ-VAE) [52] to eliminate the strong Gaussian assumption, it only conduct in single-omics data. Solely relying on the data proximity in DGMs is insufficient for learning biological complexity, e.g., inter-omics inference [53].

*2.2.2 Contrastive Learning Method.* Contrastive learning, another type of unsupervised learning method, enables the direct learning of data representations through a supervised-like task instead of approximating data distributions [54]. Very recently, Yang *et al.,* proposed a novel contrastive learning framework for comprehensive transcriptomics data representation learning [6]. Their results demonstrate superior performance compared to conventional and DGMs-based methods.

- **Limitations.** However, the work of [6] primarily attempts a single-omics data. Contrastive understanding encounters a longstanding challenge when faced with multi-omics input due to the unpredictable dimensionality. Therefore, the extension of contrastive learning to multi-omics data integration and its effectiveness in improving subtyping outcomes is still an open question.

## 2.3 Learning Biological Network

Modeling or learning biological networks is a fundamental task in systems biology. Once successful, researchers can analyze natural behavior from a statistical or more observed perspective. Classic methods, such as probabilistic Bayesian networksm have been used to the reconstruct the gene networks [55] or the interactions network of various genes and proteins for cancer[56]. Some attempts use machine learning methods to project the construction of gene regulatory networks (GRNs). The work [57] predicts GRNs occurring during seed development in Arabidopsis based on a support vector machine (SVM). More recently the work [58] uses convolutional neural networks (CNNs) to predict lung cancer from a protein-protein interactions network integrated with gene expression data. The work [59] leverages their graph representation structure using a graph neural network (GNN) to predict metabolic pathways' dynamical properties. In summary, while recent applications of biological considerations across various tasks have shown potential and promise, the field is still in its infancy for multi-omics data.

## 3 PROBLEM FORMULATION

### 3.1 Multi-Omics Data

Multi-omics data have different modalities, study emphases, and require specific high-throughput collection methods [21–23]. The fundamental entity in multi-omics data is the individual patient, and each patient comprises multiple 'omics' categories, each containing tens of thousands of measured features. Let $X_{omics} = \{X^{(i)}\}_{i=1}^{N}$ denote a longitudinal multi-omics data of $N$ patients.

Table 1: Notations used in MoCLIM

| Notation | Description |
| --- | --- |
| $X$ | Patient sample |
| $M \in \mathbb{R}^{1 \times d}$ | Multi-omics category |
| $Y$ | Cancer subtype ground truth |
| $I_A$ | Inter-omics inference |
| $M_A, \hat{M}_A$ | Anchor omics, other non-anchor omics |
| $Z$ | Latent feature representation of $X$ |
| $f_\theta(\cdot)$ | Proposed MoCLIM model |
| $\mathbf{E}_{VQ}$ | Array of features encoders |
| $\mathcal{P}$ | latent vector in categorical distribution |
| $f_{CL}(\cdot)$ | Integration function |

The $i$-th patient can be represented as a sequence of omics observations $X^{(i)} = [X_1^{(i)}, X_2^{(i)}, \cdots, X_M^{(i)}]$. The $M, M \in \mathbb{R}^{1 \times d}$ denotes the omics category with $d$ feature dimension for $i$-th patient. Let $Y = \{y_1, y_2, \cdots, y_{|Y|}\}$ denotes the set of subtypes for given a cancer. Hence, each $X^{(i)}$ is associated with a subtype label $y_{|Y|}$.

### 3.2 Multi-Omics Inference Modeling

A hope in computational biology is to model the inferential relationships or networks among various omics types to reduce the data complexity and result in more interpretable and systematic analysis [25]. This paper models the following observation in MoCLIM and tries to facilitate feature representation of distinct cancer subtypes.

OBSERVATION. *Genome-wide transcriptomics analysis is the mainstay of omics studies. Integrative analyses that evaluate cancer transcriptomics data in the context of other omics data sources often extract deeper biological insight from the data [26, 60].*

Formally, we formulate the inter-omics inference to $I_A = \{M_A, \hat{M}_A\}$. The $I_A$ can be interpreted as a network that indicates the inference relationships in multi-omics. Following the above observation, the transcriptomics category $M_A$ is established as the anchor of this inference and $\hat{M}_A$ is the other omics category. $\hat{M}_A$ as the complementary of $M_A$ that enhances the representation of cancer subtypes.

### 3.3 Problem Setting

Given a patient sample $X$, we aim to learn a lower-dimensional feature representation $Z$ that encapsulates the necessary information across all omics categories $M$. That is, $Z$ and $X$ should be equivalent for the corresponding subtype $Y$. In practice, $Z$ is produced by $Z = f_\theta(X; I_A)$ conditioned on the multi-omics inference modeling $I_A$ where $\theta$ is the model parameter. Our goal is to training a well-performed model $f_\theta(\cdot)$ (i.e., MoCLIM) with three expectations:

EXPECTATION 1. *Encode high-quality $Z$ from $X$, avoiding data issues related to high-dimensionality, scarcity, and redundancy;*

EXPECTATION 2. *Embed $I_A$ as a computational element of $f_\theta$;*

EXPECTATION 3. *Discriminate differet cancer subtypes $Y$ precisely using $Z$, which also can support subsequent biomedical analysis.*

These expectations aim to answer the *RQ1-3* described in Section 1.

## 4 MOCLIM FRAMEWORK

In this section, we present the workflow of MoCLIM, depicted in Figure 2. MoCLIM is designed in an unsupervised learning manner, as the omics data issue is scarce, high-dimensional, and noisy, and the labels are often controversial. Conventional supervised learning paradigms may not be readily applicable, which will lead to overfitting and poor generalization in such data. As a result, we implement MoCLIM to generate an informative $\mathcal{Z}$ that can effectively cluster different subtypes. The label information $\mathcal{Y}$, serving as ground truth, is only used to evaluate the performance of MoCLIM.

To meet the *expectations* (introduced in Section 3.3), MoCLIM leverages the strengths of both DGMs and contrastive learning in its architecture design. The contrastive learning strategy incorporates the concept of multi-omics inference modeling for latent feature generation. Specifically, MoCLIM consists of three modules: (1) omics-specific feature encoders; (2) multi-omics contrastive learning; and (3) biomedical evaluation.

### 4.1 Omics-Specific Encoders

Given the high-dimensionality and scarcity of data, the effectiveness of the encoders directly impacts the computational complexity of the loss function and the subtyping performance. Related multi-omics-based methods typically concatenate the features (i.e., $\mathcal{X}_1^{(i)} \oplus \mathcal{X}_2^{(i)} \cdots, \oplus \mathcal{X}_M^{(i)}$) and use a single encoder to extract global features. Here, $\oplus$ denotes the concatenation operation. The results are not satisfactory, since they ignore the modality differences and tend to forcefully compress the multi-omics data into a uniform space. Inspired by the work of [10], we propose using the Vector Quantised Variational Auto-Encoder (VQ-VAE) as the encoder [52] in MoCLIM. Furthermore, we set up an array of VQ encoders to independently learn feature representations from different omics modalities.

VQ-VAE uses a discrete latent space to represent high-dimensional or implicitly continuous data. This discrete distribution is more flexible and can more easily model arbitrary distributions because it makes no assumptions about their shape [61]. Biologically, an interesting observation is the complexity of gene expression patterns is low [53], a simple distribution assumption might be more suitable for omics data [10, 13]. Superior results have been demonstrated when applying VQ-VAE to transcriptomics data, in comparison to (Mixture) Gaussion distribution (e.g., VAEs and GANs) [10, 13]. Formally, given a set of omics data $\mathcal{X}$, the VQ encoder is defined as:

$$\mathbf{E}_{VQ}(\mathcal{X}) := \{\mathbf{E}_{VQ}^{(1)}(\mathcal{X}_1), \mathbf{E}_{VQ}^{(2)}(\mathcal{X}_2), \cdots, \mathbf{E}_{VQ}^{(M)}(\mathcal{X}_M)\} \quad (1)$$

where $\mathbf{E}_{VQ}^{(j)}(\mathcal{X}_j)$ represents a VQ encoder specifically designed to handle the $j$-th ($j \in 1, 2, ..., M$) omics modality, $\mathcal{X}_j$, of the input patient sample $\mathcal{X}$. In this manner, $M$ multi-omics modalities are processed in parallel to obtain diverse feature spaces.

In MoCLIM, we design the VQ encoders as multi-layer perceptrons (MLPs), i.e., $\mathbf{E}_{VQ}^{(j)}(\mathcal{X}_j) \mapsto \mathbf{MLP}^{(j)}(\mathcal{X}_j; \theta_j)$, where $\theta j$ denotes the parameters of the $j$-th encoder. Consequently, the output representations of the array of VQ encoders are given as:

$$(\mathcal{Z}_1, \mathcal{Z}_2, \cdots, \mathcal{Z}_M) = \mathbf{E}_{VQ}(\mathcal{X}) \quad (2)$$

Afterward, each encoding $\mathcal{Z}_j$ are embedding into a latent discrete space, which consists of $K$ latent vectors $\mathcal{P}_{1:K}$ which defines a $K$-way categorical distribution. We denote the posterior categorical distribution as $\mathcal{Z}_j^{(c)} = p(\mathcal{Z}_j = C|\mathcal{X}_j)$ as a one-hot encoding by nearest neighbor search [52]:

$$p(\mathcal{Z}_j = C|\mathcal{X}_j) = \begin{cases} 1, & \text{if } \arg\min \|\mathcal{Z}_j - \mathcal{P}_C\|_2 \\ 0, & \text{otherwise} \end{cases} \quad (3)$$

VQ encoder optimizes the conventional reconstruction loss with additional terms that drag the the latent variables towards $\mathcal{P}$:

$$\mathcal{L}_{VQ}^{(j)} := \|\mathcal{X}^{(j)} - \tilde{\mathcal{X}}^{(j)}\|_2^2 + \|\mathsf{Sg}(\mathcal{Z}) - \mathcal{P}\|_2^2 + \|\mathcal{Z} - \mathsf{Sg}(\mathcal{P})\|_2^2, \quad (4)$$

where $\tilde{\mathcal{X}}^{(j)}$ denotes the reconstructed input and $\mathsf{Sg}(\cdot)$ denotes the stop gradient operator. As a result, the total loss is the sum of the losses of all omics modalities:

$$\mathcal{L}_{VQ} = \sum_{j=1}^{M} \mathcal{L}_{VQ}^{(j)} \quad (5)$$

After extracting omics-specific representations in parallel, the resulting latent features $\mathcal{Z}$ are then passed to the subsequent integration operation of multi-omics contrastive learning.

### 4.2 Multi-Omics Contrastive Learning

Integration of multi-omics data requires the establishment of a function that maps latent features learned from different omics sources to a shared feature space. MoCLIM introduces the contrastive learning approach to conduct this integration function, $f_{CL}(\cdot)$, for further optimization of the learned latent features $\mathcal{Z}$. Contrastive learning involves learning an embedding that distinguishes between samples drawn from different feature spaces or distributions, thereby fulfilling the goal of $f_{CL}(\cdot)$. This method consists of two key components: (i) the pretext task, and (ii) the contrastive loss.

*4.2.1 Pretext Task.* Contrastive learning is an unsupervised learning, hence, the pretext refers to setting up a contrasting task to implement a form of supervised-like learning. The contrasting objects could be local-global features [62], different data modalities [63, 64], or past-future time steps [65]. This contrasting is formulated between similar and dissimilar pairs that aims to bring similar samples closer together and pushing dissimilar samples apart [54].

In MoCLIM, we set up the contrasting objects using different $\mathcal{Z}_j$ of multi-omics modalities. Let us consider $\mathcal{Z}_1^{(i)}$ and $\mathcal{Z}_2^{(i)}$ as two omics of give one patient $i$. The similar pairs, also known as *positive samples*, consist of samples from the joint distribution $\mathcal{Z}^{(+)} \sim p(\mathcal{Z}_1^{(i)}, \mathcal{Z}_2^{(i)})$. Conversely, the dissimilar pairs, known as *negative samples*, consist of samples from the product of marginals $\mathcal{Z}^{(-)} \sim p(\mathcal{Z}_1^{(i)})p(\mathcal{Z}_2^{(\tilde{i})})$, where $\tilde{i}$ denotes other patients.

In the implementation, we construct a $N$ patients mini-batch set. For each patient, we select the different omics representations $\mathcal{Z}_j$ as the positive samples and randomly select the omics representations from the other $N - 1$ patients as the negative samples.

*4.2.2 Contrastive Loss.* During the implementation of contrastive learning, the model learns to measure sample similarity by different metrics, e.g., mutual information [65], optimal transportation algorithms [66], etc. MoCLIM follows the line of mutual information as the metric, which encourages the positive-pair samples $\mathcal{Z}^{(+)}$ to have a higher mutual information than the negative-pair $\mathcal{Z}^{(-)}$.

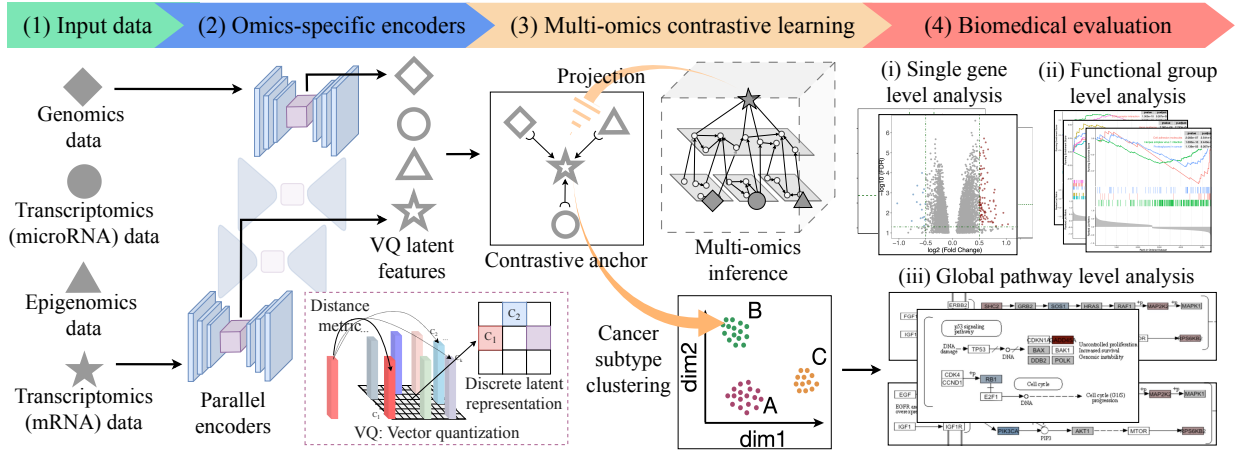

Figure 2: An overview of the MoCLIM workflow: (1) MoCLIM takes multi-omics data as input. (2) Omics-specific encoders parallelly learn latent features for each omics source. (3) Multi-omics contrastive learning with a contrastive anchor integrates the learned features. The clustering is implemented on the integrated feature. (4) Comprehensive biomedical evaluations following the feature learning help users to understand the results generated by MoCLIM.

In practice, estimating mutual information can be computationally challenging, particularly for high-dimensional and continuous data [67]. MoCLIM follows a common approximation method of the Information Noise Contrastive Estimation (InfoNCE) loss, i.e.,

$$\mathcal{L}_{\text{InfoNCE}} := \mathbb{E}_{\mathcal{X}} \left[ -\log \frac{\exp\left(\mathcal{Z} \cdot \mathcal{Z}^{(+)}/\tau\right)}{\sum_n \exp\left(\mathcal{Z} \cdot \mathcal{Z}^{(-)}/\tau\right)} \right] \quad (6)$$

where $\tau > 0$ is the temperature parameter. Minimizing Eq.(6) involves distinguishing the target $\mathcal{Z}$ from a set of $n$ ($n \leq N - 1$) randomly sampled *negative* feature vectors from the mini-batch set $N$. This approach encourages MoCLIM to extract the integrative features by maximizing the similarity among their representations.

### 4.3 Transcriptomics as Contrastive Anchor

As we have emphasized in the section 3.3, there is a common observation that mRNA transcriptomics plays a fundamental role in capturing the molecular characteristics of cancer subtypes. We hence model the inference network $\mathcal{I}_A$ into the contrastive learning of MoCLIM. That is, for $i$-th patient, $f_{CL} : \{\mathcal{Z}^{(i)}, \mathcal{Z}^{(+)}, \mathcal{Z}^{(-)}\} \to \{M_A^{(i)}, \hat{M}_A^{(i)}, M^{(\tilde{i})}\}$.

Inspired by the work of [68], given a collection of latent features $\{\mathcal{Z}_1^{(i)}, \ldots, \mathcal{Z}_M^{(i)}\}$ learned from $M$ omics categories, the transcriptomics feature ($\mathcal{Z}_{Trans}^{(i)}$) sets apart the contrasting anchor ($M_A^{(i)}$) to optimize over. Then we builds pair-wise representations between $M_A^{(i)}$ and each other view $\mathcal{Z}_{Others}^{(i)}$ by

$$\mathcal{L}_{\mathcal{I}_A} = \sum_{}^{M-1} \mathcal{L}_{\text{InfoNCE}}(\mathcal{Z}_{transcriptomics}, \mathcal{Z}_{others}) \quad (7)$$

In the implementation of contrastive learning (Eq. 6), the positive pairs are defined as the same $i$-th patient's latent features from different omics categories, i.e., $(M_A^{(i)}, \hat{M}_A^{(i)})$. The negative pairs are defined as all omics features from other patients, that is, $(M_A^{(i)}, M^{(\tilde{i})})$.

### 4.4 Training and Loss Function

The MoCLIM objective function is final defined as a weighted sum of the loss of omics-specific encoders Eq.(4) and multi-omics contrastive learning loss Eq.(7), which is formulated as follows:

$$\mathcal{L}_{\text{MoCLIM}} := \alpha \mathcal{L}_{\text{VQ}} + \sum^{M-1} \beta \mathcal{L}_{\mathcal{I}_A}, \quad (8)$$

where $\alpha, \beta$ are coefficients that are used to balance the contribution of each term. $\beta = [\beta_1, \beta_2, \cdots, \beta_{M-1}]$ can further control the contribution of each pair-wise omics representation.

### 4.5 Biomedical Evaluation Module

Generating interpretable results is essential in analyzing multi-omics data within biomedical contexts. We expect MoCLIM not only as a powerful tool for multi-omics data integration but have the potential for real-world applications within clinical settings. Therefore, the final module of the MoCLIM workflow includes a comprehensive biomedical evaluation pipeline (in Section 5.4) that follows the feature learning $\mathcal{Z}$. This pipeline incorporates a set of authoritative cancer analysis tools (i.e., a white box approach) to assist users in understanding the results generated by MoCLIM.

## 5 EXPERIMENTS

### 5.1 Datasets

To conduct a comprehensive evaluation and comparison, we constructed multi-omics datasets encompassing six representative cancer types: Breast invasive carcinoma (BRCA), Glioblastoma multiforme (GBM), Stomach adenocarcinoma (STAD), Brain Lower Grade Glioma(LGG), Uterine Corpus Endometrial Carcinoma (UCEC), and Ovarian serous cystadenocarcinoma (OV). The datasets were collected from The Cancer Genome Atlas (TCGA) [69], obtained through the world's largest cancer gene information database Genomic Data Commons (GDC) portal [70]. All patient samples were generated across various experiment platforms from cancer samples prior to treatment. Each sample consists of four categories of omics data: genomics (*gen*), microRNA transcriptomics (*mir*), epigenomics (*epi*), and mRNA transcriptomics (*mrna*). Table 2 describes the details of all experimental datasets. The details of data collection and preprocessing are described in the Appendix.

Table 2: Descriptions of the six cancer datasets

| Cancer type | Sample size | Preprocessed feature size (*gen, mir, epi, mrna*) |
| --- | --- | --- |
| BRCA | 671 | (19568, 368, 19049, 18206) |
| GBM | 414 | (19534, 453, 19053, 17455) |
| STAD | 573 | (19551, 634, 19055, 18616) |
| LGG | 434 | (19552, 443, 19062, 16245) |
| UCEC | 295 | (19546, 345, 19058, 15216) |
| OV | 302 | (19535, 547, 19054, 17226) |

### 5.2 Baseline Methods

We evaluate the performance of MoCLIM by comparing it with several baseline methods belonging to three categories:
- **Conventional methods.** SNF [36], NEMO [37], CIMLR [38], iClusterBayes [39], moCluster [40], cNMF [41]. As we introduced in Section 2, these methods are widely used in multi-omics data integration as well as cancer subtyping tasks.
- **Deep methods (1): DGMs.** We evaluate three DGMs methods. Subtype-GAN [18], which employs GAN modules to extract features from each omics data and then incorporate a shared integration layer. GMVAE [71], which employs Gaussian mixture variational autoencoders to extract features of each omics data. NVAE [72], which employs depth-wise separable Gaussian distribution for omics-specific feature extraction.
- **Deep methods (2): contrastive Learning.** To our knowledge, there is a limited number of contrastive learning methods specifically designed for multi-omics data integration and cancer subtyping studies. We conduct six contrastive learning methods to evaluate the effectiveness of our contrastive learning strategy: SCL (CPC) [65], SimCL [73], SupCL [74], MoCO [54], SWAV [66], and MFLVC [75]. The detailed descriptions can be found in the Appendix.

### 5.3 Ablation Studies

To verify the effectiveness of the core concepts of MoCLIM, we conduct the following three ablation studies:
- **The anchor omics ablation.** We select different omics data as the anchor omics, assessing the suitability of transcriptomics.
- **The number of omics ablation.** We progressively exclude omics data sources from the input to evaluate whether MoCLIM can maintain performance even as the input complexity increases.
- **Contrastive learning strategy ablation.** We remove the contrastive learning and use another strategy for MoCLIM evaluation.

### 5.4 Biomedical Analysis Pipeline

We conduct three phases of biomedical cancer analysis as the sample to further evaluate the utilization and effectiveness of MoCLIM: single gene level, functional group level, and global pathway level.

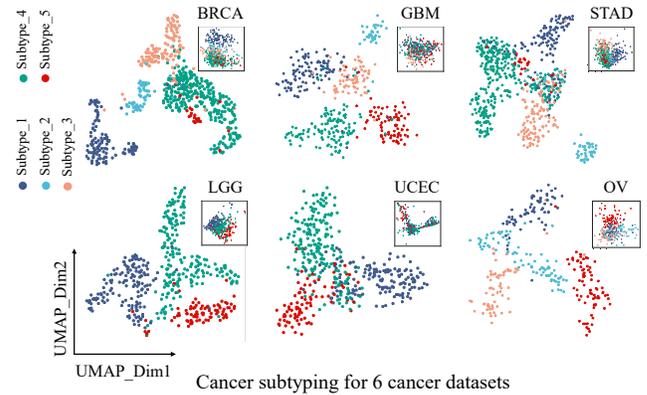

Figure 3: UMAP visualization illustrates the subtyping results across six distinct cancer types, using the latent features learned from the MoCLIM. Each patient case is marked with colors representing its corresponding subtype. In comparison, the upper-right box shows the subtyping results obtained from the original multi-omics data which ignores the modality differences.

Since single gene level analysis is an initial step for cancer subtype analysis, we does not show the results in main body of the paper, the analyasis results are shown in the Appendix.
- **Fuctional group level: gene set enrichment analysis (GSEA).** We further conduct the gene set enrichment analysis (GSEA), which determines the between-cluster functional differences by identifying enriched gene sets associated with each subtype. The GSEA analysis uses the pre-ranked approach, where genes are ranked based on their differential expression between pairwise groups. The log2 fold change of gene abundance is the ranking metric, capturing the magnitude of expression changes. We use GSEA to evaluate the effectiveness of representation comparing with the raw data while serving for 'contrastive learning strategy ablation study'.
- **Global pathway level: cancer-related pathway.** In the third phase, we conduct a global pathway-level analysis specific to cancer-related pathways. This tool can evaluate the differences in resulted clusters. Notice that the last analysis phase can be highly flexible. Building upon the information obtained from the previous two-phase, we can choose and focus on specific pathways that are highly related to the cancer we are interested in. In the main body, we map the related gene expression information of patients from two resulting subtype clusters to the chosen pathway map. The whole pathway map is shown in the Appendix.

### 5.5 Evaluation Metrics and System Settings

MoCLIM is an unsupervised representation learning framework, we conduct clustering task incorporating with the patient label to evaluate the effectiveness and performance of cancer subtyping. We employed three widely adopted evaluation metrics utilized in clustering studies: the silhouette coefficient (SIL), normalized mutual information (NMI), and adjusted rand index (ARI).

We implemented all experiments using the Pytorch v1.4 framework and conducted them on a server with an NVIDIA GeForce RTX 3090Ti GPU. The medical evaluations are conducted on R studio. The model architecture, architecture ablation study, and hyperparameter fine tuning can be found in the Appendix.

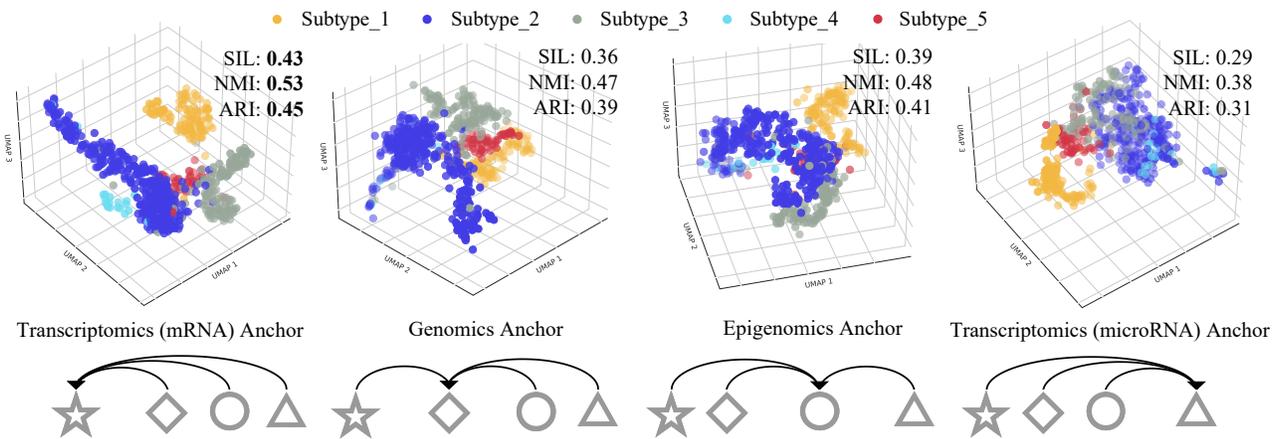

Figure 4: The results of the BRCA subtyping task using different omics data as the anchor for contrastive learning. The mRNA transcriptomics anchor outperforms other omics anchors: it shows the most well-separated subtyping results and is the only one that accurately identifies subtype 4.

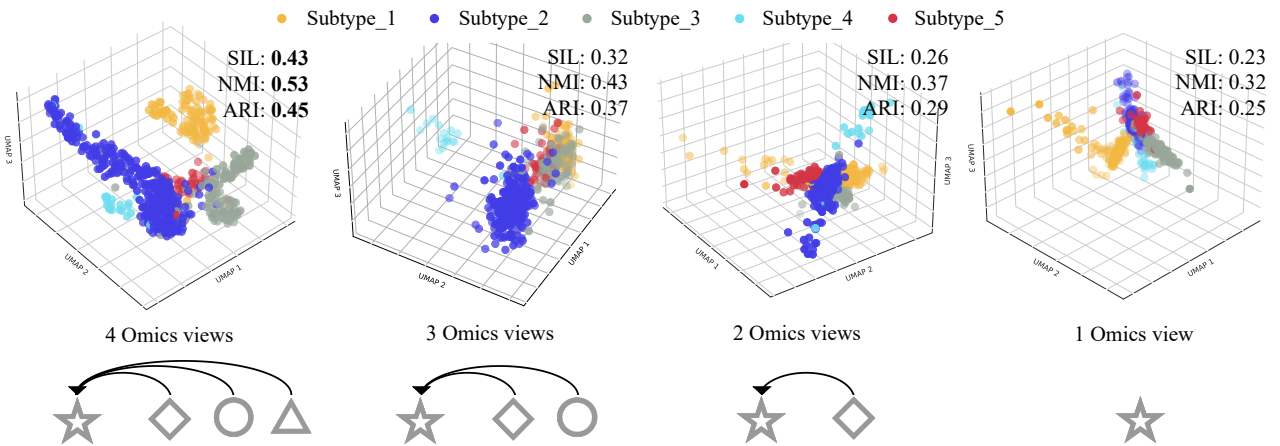

Figure 5: The results of the BRCA subtyping task where other omics data sources are progressively excluded from the input after making mRNA transcriptomics the anchor. As each omics data source is removed, there shows a consistent decrease in model performance.

### 5.6 Results

*5.6.1 Overall Cancer Subtyping Performance.* Figure. 3 shows the UMAP visualized subtyping results obtained through learned features in MoCLIM across six tasks of different cancer types. These results are compared with the ones obtained through aligned original multi-omics data. Different cancer subtypes can be well-separated in the learned latent feature space, with clear boundaries between groups. These observations are consistent across all six cancer subtyping tasks. In each subtype (cluster), patient samples display noticeably tighter distributions in the learned feature spaces compared to the original multi-omics data space. Importantly, our experiments clearly show the importance of using the suitable omics-specific learning and array VQ encoders for cancer subtyping tasks. In most cases, aligning multi-omics data at the sample level ignores the modality differences and leads to unexpectable dimensionality of each input. It cannot learn the effective representation between distinct subtype groups. The clustering visualization of one sample cancer (BRCA) can be found in the left of Figure. 4 and Figure. 5. Notably, these two figures only show results for BRCA cancer.

*5.6.2 Anchor Omics Ablation Results.* Figure 4 illustrates the subtyping results of using different omics data as the anchor for contrastive learning. These results are visualized in a three-dimensional feature space using UMAP. Specifically, we iteratively select mRNA transcriptomics, genomics, epigenomics, and miRNA transcriptomics as the contrasting anchor to learning an anchor-omics feature representation. The result demonstrates that mRNA Transcriptomics yields the most well-separated subtyping results and outperforms other omics data sources. The subtyping results achieve a SIL of 0.43, NMI of 0.53, and ARI of 0.45, which are higher than the second-best result by epigenomics (SIL of 0.39, NMI of 0.48, and ARI of 0.41). Furthermore, modeling transcriptomics as the anchor can accurately identify subtype 4, which challenges to other ablations. This outcome aligns with our initial idea that mRNA Transcriptomics data is the most appropriate choice for guiding the integration of multi-omics data. MoCLIM can effectively model this consideration and omics inference.

Table 3: Baseling comparison results on Silhouette (SIL), Normalized Mutual Information (NMI), and Adjusted Rand Index (ARI) for the proposed MoCLIM, related conventional methods, and deep-based methods. Bold denotes the best results.

| Method | BRCA | | | LGG | | | GBM | | | UCEC | | | STAD | | | OV | | |
|---|---|---|---|---|---|---|---|---|---|---|---|---|---|---|---|---|---|---|
| | SIL | NMI | ARI | SIL | NMI | ARI | SIL | NMI | ARI | SIL | NMI | ARI | SIL | NMI | ARI | SIL | NMI | ARI |
| Conventional Methods | | | | | | | | | | | | | | | | | | |
| SNF | 0.19 | 0.32 | 0.23 | 0.17 | 0.32 | 0.21 | 0.18 | 0.31 | 0.20 | 0.15 | 0.29 | 0.21 | 0.17 | 0.32 | 0.22 | 0.14 | 0.28 | 0.21 |
| NEMO | 0.16 | 0.33 | 0.25 | 0.15 | 0.31 | 0.22 | 0.16 | 0.31 | 0.22 | 0.12 | 0.28 | 0.21 | 0.15 | 0.32 | 0.22 | 0.11 | 0.25 | 0.21 |
| CIMLR | 0.05 | 0.22 | 0.20 | 0.05 | 0.21 | 0.20 | 0.04 | 0.20 | 0.19 | 0.04 | 0.17 | 0.15 | 0.06 | 0.22 | 0.23 | 0.03 | 0.15 | 0.13 |
| iClusterBayes | 0.14 | 0.22 | 0.18 | 0.14 | 0.22 | 0.18 | 0.14 | 0.21 | 0.15 | 0.13 | 0.19 | 0.15 | 0.16 | 0.22 | 0.15 | 0.12 | 0.19 | 0.15 |
| moCluster | 0.04 | 0.15 | 0.14 | 0.04 | 0.15 | 0.13 | 0.04 | 0.13 | 0.13 | 0.03 | 0.12 | 0.12 | 0.05 | 0.17 | 0.15 | 0.03 | 0.11 | 0.09 |
| cNMF | 0.04 | 0.19 | 0.18 | 0.03 | 0.18 | 0.17 | 0.04 | 0.17 | 0.15 | 0.03 | 0.14 | 0.12 | 0.04 | 0.19 | 0.17 | 0.03 | 0.12 | 0.11 |
| Deep-based Methods | | | | | | | | | | | | | | | | | | |
| GAN-Subtype | 0.14 | 0.21 | 0.15 | 0.14 | 0.21 | 0.15 | 0.13 | 0.20 | 0.16 | 0.12 | 0.19 | 0.15 | 0.15 | 0.23 | 0.17 | 0.12 | 0.18 | 0.15 |
| GMVAE | 0.02 | 0.14 | 0.12 | 0.02 | 0.14 | 0.13 | 0.02 | 0.12 | 0.13 | 0.02 | 0.13 | 0.12 | 0.02 | 0.13 | 0.14 | 0.02 | 0.12 | 0.12 |
| NVAE | 0.03 | 0.13 | 0.12 | 0.03 | 0.13 | 0.13 | 0.02 | 0.11 | 0.12 | 0.03 | 0.11 | 0.09 | 0.03 | 0.14 | 0.14 | 0.03 | 0.09 | 0.06 |
| SCL (CPC) | 0.21 | 0.31 | 0.24 | 0.19 | 0.31 | 0.22 | 0.18 | 0.32 | 0.22 | 0.16 | 0.27 | 0.21 | 0.19 | 0.33 | 0.22 | 0.15 | 0.25 | 0.19 |
| SimCL | 0.25 | 0.37 | 0.28 | 0.23 | 0.34 | 0.27 | 0.21 | 0.35 | 0.23 | 0.19 | 0.26 | 0.21 | 0.24 | 0.36 | 0.29 | 0.18 | 0.24 | 0.19 |
| SupCL | 0.23 | 0.34 | 0.25 | 0.23 | 0.31 | 0.24 | 0.21 | 0.32 | 0.21 | 0.17 | 0.29 | 0.19 | 0.24 | 0.32 | 0.23 | 0.16 | 0.26 | 0.17 |
| MoCO | 0.24 | 0.35 | 0.25 | 0.24 | 0.34 | 0.25 | 0.21 | 0.33 | 0.21 | 0.18 | 0.27 | 0.20 | 0.25 | 0.35 | 0.26 | 0.16 | 0.25 | 0.19 |
| SWAV | 0.32 | 0.41 | 0.35 | 0.31 | 0.38 | 0.31 | 0.30 | 0.35 | 0.27 | 0.28 | 0.35 | 0.29 | 0.30 | 0.37 | 0.32 | 0.25 | 0.33 | 0.26 |
| MFLVC | 0.21 | 0.32 | 0.23 | 0.18 | 0.29 | 0.23 | 0.18 | 0.31 | 0.21 | 0.15 | 0.25 | 0.20 | 0.17 | 0.29 | 0.23 | 0.14 | 0.23 | 0.19 |
| MoCLIM | **0.43** | **0.53** | **0.45** | **0.40** | **0.51** | **0.43** | **0.41** | **0.51** | **0.42** | **0.39** | **0.48** | **0.38** | **0.41** | **0.52** | **0.43** | **0.35** | **0.46** | **0.35** |

*5.6.3 The Number of Omics Ablation Results.* Figure 5 presents another set of subtyping results obtained when progressively excluding other omics data sources from the input after making mRNA transcriptomics as the anchor in MoCLIM. The results clearly indicate a consistent decrease in model performance as each omics data source is removed. The SIL decreases from 0.43 for the four omics inputs to 0.32, 0.26, and 0.23 for three, two, and only one omics input, respectively. This finding highlights the differential contributions of different omics studies to cancer subtyping tasks again. It further indicates that by explicitly modeling the modality differences, each omics data source can offer unique and valuable information that enhances the overall subtyping performance. The decreasing trend in model performance further proves the necessity of using multi-omics to complementary to finding the gene expression changes for cancer subtyping. With increasing the omics input, MoCLIM avoid the risk of noise increase, and can effectively captures the synergistic effects for accurately identifying subtypes.

*5.6.4 Baseline Comparison Results.* Table 3 presents the comparison results between the proposed MoCLIM and other baseline works. MoCLIM outperforms other methods in all three metrics for cancer subtyping across six datasets. In particular, for smaller datasets such as LGG, UCEC, and OV (comprising only 434, 295, and

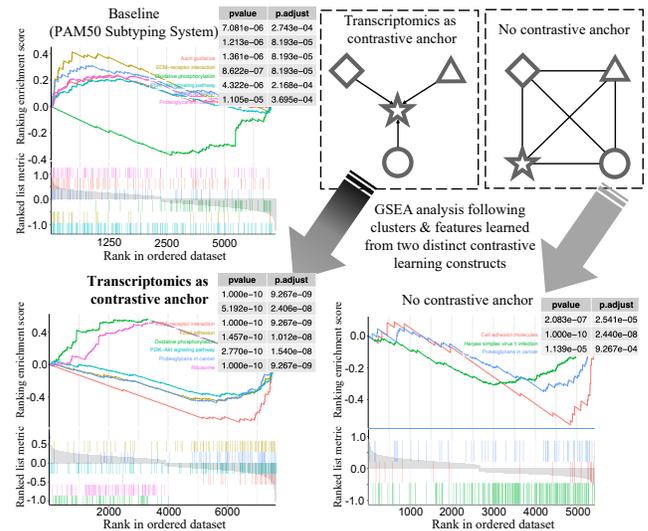

Figure 6: GSEA analysis results of the BRCA subtyping task. The figures correspond to the subtypes identified with the PAM50 baseline, transcriptomics as the contrastive anchor (MoCLIM), and no contrastive anchor. Curves in different colors with corresponding P-values indicate the functional groups in which the genes are enriched.

302 samples, respectively), MoCLIM exhibits robust performance, achieving 0.48, 0.52, and 0.46 in NMI, respectively. These values are 0.13, 0.16, and 0.21, higher than the second-best result. Deep learning-based methods, specifically those utilizing contrastive learning, consistently outperform conventional methods. Notably, two VAE-based methods, GMVAE and NVAE, which make complex data distribution assumptions, show poor performance across all datasets. This indicates the single categorical distribution assumed in the MoCLIM VQ encoder is more suitable for omics data.

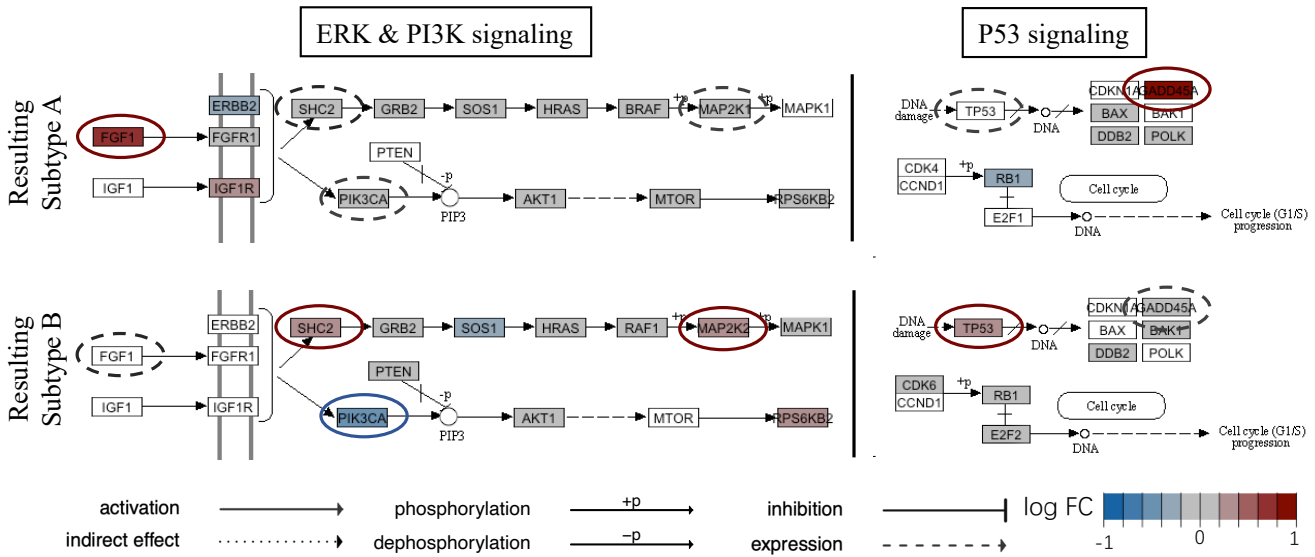

**Figure 7: Visualization of the ERK&PI3K signaling pathway and the P53 signaling pathway in two resulting BRCA subtypes. Red circles and blue circles denote genes significantly upregulated and downregulated in the pathway. Dashed gray circles in comparison denote no change.**

*5.6.5 Biological Evaluation Results.* Fig. 6 is the GSEA analysis results in the BRCA subtyping task. Curves in different colors indicate the different functional groups in which the genes are enriched. More curves observed in the GSEA result indicate a more distinct bio-functional characteristic of the identified subtype; a smaller P-value also indicates a better-identified enriched group. We present three results: the figures sit in upper-left, lower-left, and lower-right corresponding to the subtypes identified with the baseline PAM50 subtyping system, transcriptomics as the contrastive anchor (modeling the omics-inference), and no contrastive anchor, respectively. Compared to the other two methods, MoCLIM achieves the best results with more enriched functional groups and significantly smaller P-values. In particular, the MoCLIM result shows notable enrichment of gene sets related to ECM-receptor interaction and the PI3K-Akt signaling pathway, distinguishing it from the baseline method. However, the result without a contrastive anchor is even worse than the baseline method. This observation indicates that in addition to the omics-specific feature encoders, a transcriptomics contrastive anchor is also crucial for the final learned features: without an ideal contrastive anchor, incorporating more omics data will only introduce more noise and destroy the feature learning process.

Fig. 7 shows parts of the cancer-related pathway analysis results, which focused on two representative pathways in BRCA: the ERK & PI3K signaling pathway and the P53 signaling pathway. The whole resulting pathway map can be found in the Appendix. The upper and lower-half figure shows two subtypes' gene active situation in two pathways. The gene in the red and blue circles indicate significant upregulated and downregulated in the pathway, compared with the ones in gray. This difference may implicate this biomarker as well as a potential therapeutic target. For example in subtype A, we identified a notable over-expression of FGF1, a gene known to play a crucial role in the regulation of cell growth and proliferation [76]. In contrast, subtype B did not exhibit a pronounced expression of FGF1. This suggests that targeting FGF1 may offer novel therapeutic opportunities for subtype one patients.

## 5.7 Conclusion

This paper introduced MoCLIM, a multi-omics contrastive learning framework for accurate cancer subtype identification. MoCLIM treats multi-omics integration as multi-modal learning and independently extracted representations from each type of omics data. These representations are then combined using a contrastive learning framework anchored in transcriptomics. This contrasting approach can be interpreted as a projection of inter-omics inference, as observed in biological networks. As a result, MoCLIM is general and flexible, capable of integrating any omics data without amplifying the risk of noise or introducing redundant information. Importantly, we showed the extension of MoCLIM to a series of biomedical evaluations, e.g., pathway analysis. We hope MoCLIM is a valuable tool for multi-omics-related research or other biomedical applications. We leave the in-lab experiments in our future work to validate its practical effectiveness.

## 5.8 Acknowledgments

This work was partially supported by JSPS KAKENHI Grant-in-Aid for Scientific Research Number JP21H03446, NICT 03501, and JST-AIP JPMJCR21U4.

# Appendix of "MoCLIM: Towards Accurate Cancer Subtyping via Multi-Omics Contrastive Learning with Omics-Inference Modeling"

## 1 DATASETS PREPROCESSING

**Transcriptomics (mRNA&miRNA) data.**: Inconsistencies in gene annotations often result in the absence of certain expression features across different platforms. To ensure platform independence, we initially removed the cross-platform lost features. For the transcriptomics data generated from the Hi-Seq platform, we converted the scaled estimates in the original gene-level RSEM (RNA-Seq by expectation maximization) files to FPKM (fragments per kilomillion base) mapped reads data. For the remaining data generated from the Illumina GA and Agilent array platforms, we initially identified and removed all non-human expression features. Subsequently, we applied a logarithmic transformation to the converted data. To eliminate potential noise, we identified and eliminated features with zero expression levels (based on a threshold of more than 10% of samples) or missing values (designated as N/A). The missing data imputation was performed at last using the R package IMPUTE [1] to ensure a more complete dataset.

**Genomic & epigenomic data.**: For the genomic level and epigenomic level, we focused on copy number variations (CNVs) data and DNA methylation data, respectively. Quality control checks were first applied to ensure data integrity for the original data collected (referred to as level-3 data). To account for systematic biases and technical variations across the samples, a median-centering normalization was implemented using the R package limma [2]. To capture only somatic mutations and exclude germline mutations, a filtering step was performed for CNV data, retaining only those mutations that originated from somatic. We used the R package GAIA [3] to identify the recurrent alterations in the cancer genome first since the level-3 data denote all aberrant regions along the genome resulting from copy number variation segmentation. We further used the R package BiomaRt [4] to annotate the aberrant recurrent genomic regions to verify the significantly amplified or deleted genes. Similar filtering and annotation steps are applied to DNA methylation data to focus on cancer-relevant genomic regions.

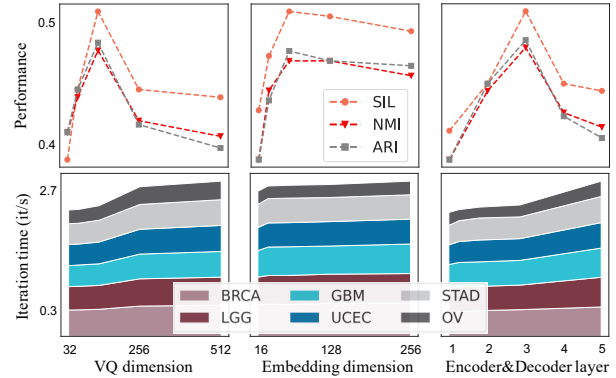

**Figure 1: The visualization of the relationship between different hyperparameter configurations and their effect on the performance of the model, accompanied by the corresponding training iteration times.**

## 2 OBSERVATION ON HYPERPARAMETERS

**Table 1: Model hyperparameter settings. The optimal settings are highlighted in bold font.**

| Parameter | Value |
| --- | --- |
| Vector quantization dimension | {32, 64, **128**, 256, 512} |
| Embedding dimension | {16, 32, **64**, 128, 256} |
| #Encoder&Decoder layer | {1, 2, **3**, 4, 5} |
| Dropout rate | {**0.2**, 0.5, 0.8} |
| #Training epoch | 200 |
| Batch size | 32 |
| #Parameters | $8.6 \times 10^6$ |
| Learning rate | $10^{-4}$ |

Figure 1 illustrates the efficiency of our model and its sensitivity to different hyperparameter selections. From the figure, we observe that the model exhibits higher sensitivity to the selection of the VQ dimension and the number of encoder and decoder layers. Changes in these hyperparameters result in noticeable variations in both the model's performance and the time required for model training. On the other hand, the embedding dimension appears to have a relatively minor effect on the model's performance, as indicated by the smaller changes observed in comparison. These findings emphasize the importance of careful consideration and tuning of hyperparameters, particularly the VQ dimension and the number of encoder and decoder layers, to achieve optimal performance and training efficiency. The detailed setting and the best combinations of the model parameters can be found in Table 1.

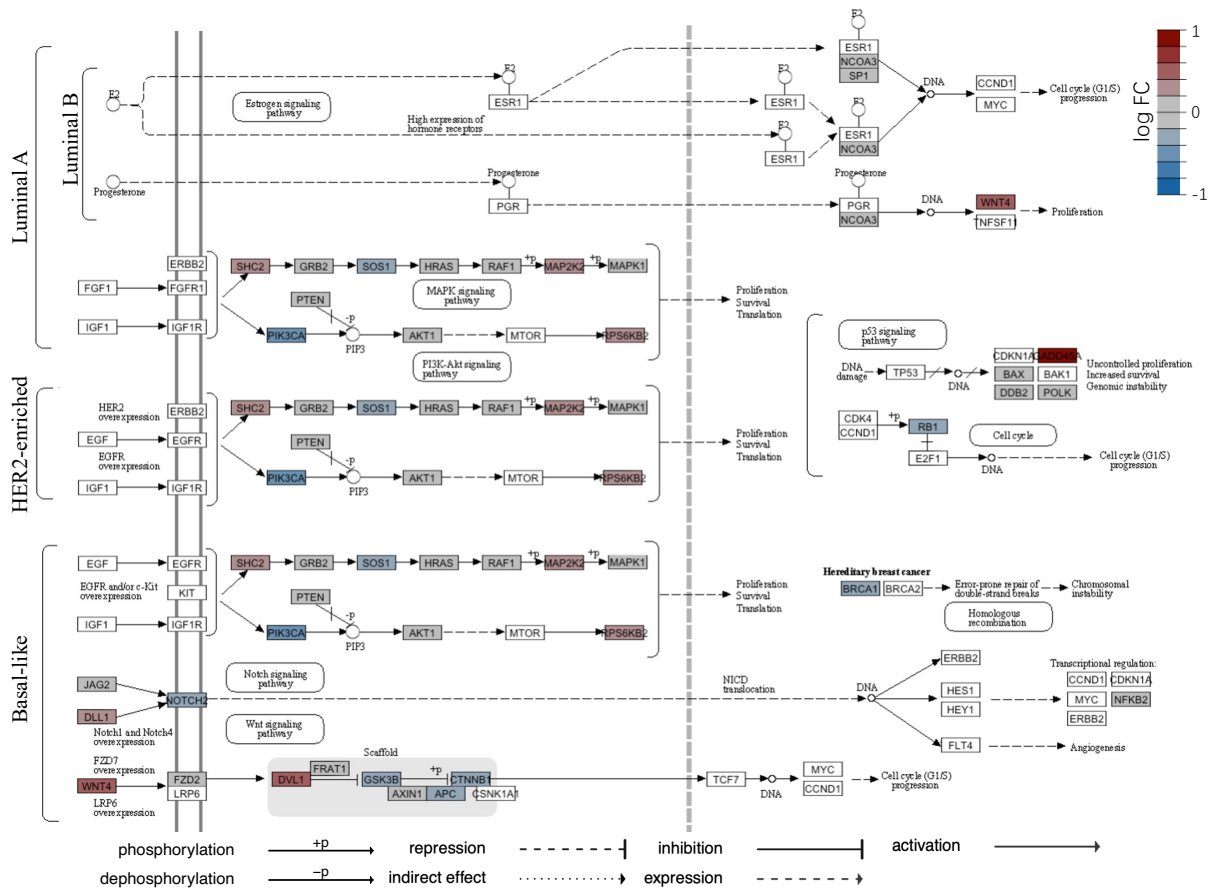

**Figure 2: Whole graph of breast cancer-related pathway**

## 3 BIOMEDICAL EVALUATION PROCESS

**Single gene level analysis.**: We calculated the log2 fold change in gene abundance between pairwise groups and determined the significance of expression changes using Student's t-test. To correct for the false discovery rate, p-values were adjusted using the Benjamini-Hochberg procedure. We considered a gene to be significant if it had an adjusted p-value less than 0.05 and a log2 fold change greater than or equal to 1.2. The resulting DEGs were categorized into up-regulated and down-regulated sets based on their fold changes and can be utilized for subsequent analysis phases.

**Fuctional group level: gene set enrichment analysis (GSEA).**: The GSEA analysis was performed using a pre-ranked approach, where genes were ranked based on their differential expression between pairwise groups. The log2 fold change of gene abundance was utilized as the ranking metric, capturing the magnitude of expression changes. To ensure robustness and reliability, the significance of gene set enrichment was rigorously assessed through permutation testing, effectively controlling for false discoveries and accounting for multiple hypothesis testing.

**Global pathway level: cancer-related pathway.**: Notice that the last analysis phase can be highly flexible. Building upon the information obtained from the previous two-phase, we can choose and focus on specific pathways that are highly related to the cancer we are interested in. Then we mapped the related gene expression information of patients from different resulting subtype clusters to the chosen pathway map.

## 4 SUPPLEMENT ON BIOLOGICAL EVALUATION RESULTS

Fig. 3 shows the volcano plot depicting the DEGs in the resulting subtype A compared to subtype B in BRCA. In this plot, up-regulated DEGs are denoted by red dots, down-regulated DEGs by blue dots, and non-significantly differentially expressed genes by gray dots.

Among the identified DEGs, several genes have been extensively reported as being associated with cancer progression. Notable examples include BRCA1, WNT4, and NOTCH2. BRCA1 is well-known for its involvement in hereditary breast cancer and plays essential roles in cell cycle regulation, DNA damage response, and transcriptional control [5]. Dysregulation of the WNT4 gene, which encodes a protein belonging to the Wnt signaling pathway, has been linked to tumor growth, invasion, and metastasis [6]. Similarly, the NOTCH2 gene, a member of the Notch receptor family, is critical in cell fate determination, development, and tissue homeostasis, and has been implicated in tumor initiation, progression, and therapy resistance [7]. These findings highlight the ability of latent features-based subtyping to identify potential biomarkers for clinically distinguishing different cancer subtypes.

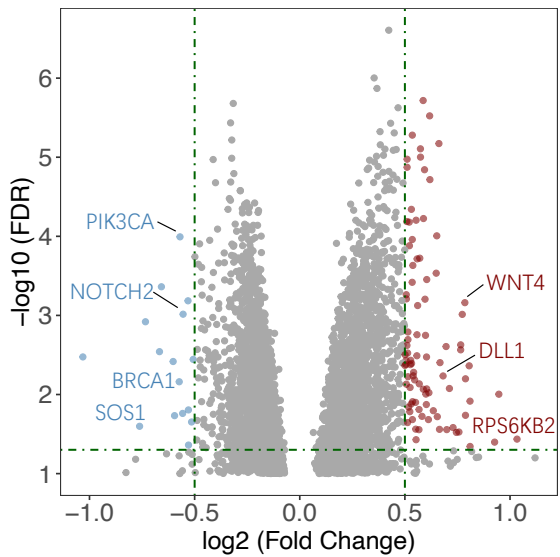

**Figure 3: Volcano plot of differential expression genes (DEGs) in cluster subtype one versus other clusters in breast cancer. Red, blue, and gray dots denote up-regulated, down-regulated, and non-significantly DEGs.**

Fig. 2 illustrated pathway map considers the regulatory relationship of key genes under specific pathways (Notch signaling pathway, MAPK signaling pathway, etc.) in subtypes Luminal A, Luminal B, HER2-enriched, and Basal-like of the PAM50 system.